\documentclass[a4paper]{article}
\usepackage{spconf,amsmath,graphicx,cite}
\usepackage[linesnumbered, ruled]{algorithm2e}
\usepackage{epstopdf}

\let\oldnl\nl
\newcommand{\nonl}{\renewcommand{\nl}{\let\nl\oldnl}}

\title{RELAXED CONCENTRATED MLE FOR ROBUST CALIBRATION OF RADIO INTERFEROMETERS}
\name{V. Ollier$^*$ $^\ddagger$,  M. N. El Korso$^\dagger$, R. Boyer$^\ddagger$, P. Larzabal$^*$ and M. Pesavento$^\mathsection$\thanks{This work was supported by the
following projects: MAGELLAN (ANR-14-CE23-0004-01) and by the iCODE institute, research project of the IDEX Paris-Saclay.}}
\address{%
    $^*$ SATIE, UMR 8029, ENS Cachan, Universit\'{e} Paris-Saclay, Cachan, France\\
    $^\dagger$ LEME, EA 4416, Universit\'{e} Paris-Ouest, Ville d'Avray, France \\
    $^\ddagger$ L2S, UMR 8506, Universit\'{e} Paris-Sud, Gif-sur-Yvette, France\\
    $^\mathsection$ Communication Systems Group,
Technische Universit\"{a}t Darmstadt, Darmstadt, Germany\\
}
\begin{document}

\maketitle
\begin{abstract}
In this paper, we investigate the calibration of radio interferometers in which Jones matrices are considered to model the interaction between the incident electromagnetic field and the antennas of each station. Specifically, perturbation effects are introduced along the signal path, leading to the conversion of the plane wave into an electric voltage by the receptor. In order to design a robust estimator, the noise is assumed to follow a spherically invariant random process (SIRP). The derived algorithm is based on an iterative relaxed concentrated maximum likelihood estimator (MLE), for which closed-form expressions are obtained for most of the unknown parameters.
\end{abstract}
\section{Introduction}
\label{sec:intro}

The new generation of radio telescopes, such as the low frequency array (LOFAR) \cite{van2013lofar} and the square kilometre array (SKA) \cite{dewdney2009square}, are characterized by a large number of receiving elements, large collecting area and dynamic range, wide field of view, high sensitivity and resolution, huge amount of measurement data, etc., which entails a certain number of scientific challenges. In radio interferometry \cite{thompson2008interferometry}, one of the most important challenges is notably the calibration \cite{wijnholds2010calibration}.

Calibration involves the estimation and the correction of different unknown perturbations introduced along the signal path, e.g., due to the environment (atmosphere, ionosphere) or the artifacts in the instruments (electronic gain, bandpass, station beam shape and orientation, sidelobe contamination, etc.). All these physical corruption effects, which may be direction dependent \cite{smirnov2011revisiting2}, are involved in the radio interferometer measurement equation and can be modeled with the help of Jones matrices \cite{hamaker1996understanding, smirnov2011revisiting}. Besides, many faint sources are present in radio interferometric data and can be considered as outliers in the calibration procedure, leading to deviations from the commonly assumed Gaussian noise model \cite{yatawatta2009radio, kazemi2011radio}.

To overcome these drawbacks, we intend to robustify the calibration scheme by using a wider distribution class than the Gaussian one, to model the noise. In doing so, we do not specify precisely the noise distribution, unlike \cite{kazemi2013robust} where the Student's t-distribution is considered, and we use a broad class of distributions gathered under the so-called spherically invariant random distribution \cite{jay2002detection, yao2003spherically}. A spherically invariant random process (SIRP) is
described as the product of a texture parameter: a positive random variable, and a speckle component: a Gaussian process, resulting in a two-scale compound-Gaussian distribution.
Under SIRP noise, the maximum likelihood (ML) method can be used to estimate the unknown parameters \cite{wang2006maximum}. In our case, to obtain closed-form expressions and to reduce the computational complexity of our problem, the ML estimates are derived in an iterative way with a sequential updating procedure\cite{zhang2015maximum}. This is the iterative concentrated ML technique. However, a numerical optimization process still needs to be performed and can be computed efficiently for instance with the Levenberg-Marquardt (LM) solver \cite{madsen1999methods, nocedal2006numerical}.
%

In this paper, we use the following notation: symbols $\left( \cdot \right) ^{T}$, $\left( \cdot \right)^{\ast }$, $\left( \cdot \right) ^{H}$ denote, respectively, the transpose, the complex conjugate and the Hermitian operator. The Kronecker product is represented by $\otimes$, $\mathrm{E}\{\cdot\}$ denotes the expectation operator and $\mathrm{bdiag}\{\cdot\}$ is the block-diagonal operator. The trace and determinant operators are, respectively, referred to by $\mathrm{tr}\left\{ \cdot \right\} $ and $ |\cdot|$. Finally, the symbol $\mathbf{I}_{B}$ represents the $B \times B$ identity matrix and $\mathrm{vec}(\cdot)$ stacks the columns of a matrix on top of one another.

\section{Data model}
\label{sec:model}

Let us consider $D$ signal sources impinging on a station of $M$ antennas.
Each electromagnetic plane wave is observed by $M$ antennas and can be decomposed as two orthogonal
polarization directions $(x,y)$. Each antenna is composed of two receptors, which are sensitive to a particular polarization \cite{hamaker1996understanding}. Thus, the measured voltage at the \textit{p}-th antenna due to the \textit{i}-th signal source impinging on it is written as \cite{hamaker1996understanding, smirnov2011revisiting}
\begin{equation}
\bar{\mathbf{v}}_{i_{p}} = \mathbf{J}_{i_{p}}(\boldsymbol{\theta}) \mathbf{s}_{i}
\end{equation}
where the relation between each incoming radiation $\mathbf{s}_{i} = [
  s_{i_{x}},s_{i_{y}}]^{T}$ and the generated voltage at each antenna $ \bar{\mathbf{v}}_{i_{p}} = [
  v_{i_{p_{x}}},
v_{i_{p_{y}}}]^{T}$ is given by a $2 \times 2$ Jones matrix $\mathbf{J}_{i_{p}}(\boldsymbol{\theta})$, parametrized by the unknown vector $\boldsymbol{\theta}$. The Jones matrix  accounts for the
different perturbations introduced on the path from the \textit{i}-th source to the \textit{p}-th sensor.
Hence, for a given source-antenna pair, we measure two output signals, i.e., one for each polarization. Since each Jones matrix is associated with a source-antenna pair, the total number of Jones matrices is $D M$.

Typically, in radio astronomy, signals collected by a pair of antennas $(p,q)$, i.e., two pairs of output signals, are correlated. Then, different crosscorrelation measurements, called visibilities, are computed for different antenna pairs, with a specific baseline.  The total number of antenna pairs  is
$B=\frac{M(M-1)}{2}$
and for a given $(p,q)$ antenna pair, the $2 \times 2$ visibility
matrix, in the noiseless case, is denoted by $\tilde{\mathbf{V}}_{pq}=\mathrm{E}\{\bar{\mathbf{v}}_{p}\bar{\mathbf{v}}_{q}^{H}\}$
and written as
\begin{equation}
\label{ME}
\tilde{\mathbf{V}}_{pq}  =
\sum_{i=1}^{D}\mathbf{J}_{i_{p}}(\boldsymbol{\theta})\mathbf{C}_{i}
\mathbf{J}_{i_{q}}^{H}(\boldsymbol{\theta}) \ \
\text{for} \ \ p<q, \ \ p,q \in [1, \ldots, M],
\end{equation}
in which, for the \textit{i}-th source, $\mathbf{C}_{i}
=\mathrm{E}\{\mathbf{s}_{i}\mathbf{s}_{i}^{H}\}$ is the $2 \times 2$ intrinsic source coherency
matrix, known from prior knowledge.
As one can notice, this equation is composed of the contributions from $D$ discrete polarized calibrator sources ($D >1$ to avoid calibration ambiguities \cite{smirnov2011revisiting2}) with uncorellated radiations and the corresponding path effect modeled by the Jones matrices.

Using \cite [p.~424]{van2013signal}, the vectorized form of (\ref{ME}) can be written as
\begin{equation}
\tilde{\mathbf{v}}_{pq}=\mathrm{vec}(\tilde{\mathbf{V}}_{pq})
=\sum_{i=1}^{D}\mathbf{u}_{i_{pq}}(\boldsymbol{\theta})
\end{equation}
where $\mathbf{u}_{i_{pq}}(\boldsymbol{\theta})=\left(\mathbf{J}^{\ast}_{i_{q}}(\boldsymbol{\theta})\otimes
\mathbf{J}_{i_{p}}(\boldsymbol{\theta})\right)\mathbf{c}_{i}$, in which $\mathbf{c}_{i}=\mathrm{vec}(\mathbf{C}_{i})$.
To consider a more realistic scenario, we introduce a noise vector $\mathbf{n}_{pq}$ for each antenna pair $(p,q)$ such that the visibility vector becomes
\begin{equation}
\mathbf{v}_{pq}=\tilde{\mathbf{v}}_{pq}+\mathbf{n}_{pq}.
\end{equation}
The full visibility vector $\mathbf{x}$ of length $4 B$ is given by
\begin{equation}
\mathbf{x} = \left[\begin{array}{c}
  \mathbf{v}_{12}\\
\mathbf{v}_{13}\\
\vdots\\
\mathbf{v}_{(M-1)M}
\end{array}\right]
=\sum_{i=1}^{D}\mathbf{u}_{i}(\boldsymbol{\theta}) +\mathbf{n}
\end{equation}
where $\mathbf{u}_{i}(\boldsymbol{\theta})=
  \left[\mathbf{u}^{T}_{i_{12}}(\boldsymbol{\theta}),
\mathbf{u}^{T}_{i_{13}}(\boldsymbol{\theta}),
\ldots,
\mathbf{u}^{T}_{i_{(M-1)M}}(\boldsymbol{\theta})\right]^{T}$ and \\ $\mathbf{n}=\left[
  \mathbf{n}^{T}_{12},
\mathbf{n}^{T}_{13},
\ldots,
\mathbf{n}^{T}_{(M-1)M}
\right]^{T}$.


\section{MAXIMUM LIKELIHOOD ESTIMATION}
\label{sec:MLestim}

The $D$ signal sources correspond to the brightest sources, while the weak ones are considered as noise. Therefore, outliers may appear and the Gaussian noise assumption may not be fulfilled \cite{lange1989robust}, e.g., a Student's t-distribution may be used \cite{kazemi2013robust}. To cope with different noise distributions, specifically non-Gaussian noise modeling, and to achieve robust calibration w.r.t.~outliers, we consider a SIRP, which is defined for each antenna pair as
\begin{equation}
\label{sirp}
\mathbf{n}_{pq} = \sqrt{\tau_{pq}} \  \mathbf{g}_{pq}
\end{equation}
where the random variable $\tau_{pq}$ is positive and real. This power factor varies independently according to the antenna pair considered and, in the radar context, is called texture. The speckle component $\mathbf{g}_{pq}$ is a complex zero-mean Gaussian process with an unknown covariance matrix $\boldsymbol{\Omega}$, i.e.,
\begin{equation}
\label{constraint}
\mathbf{g}_{pq} \sim \mathcal{CN}(\mathbf{0},\boldsymbol{\Omega})  \ \
\text{such that} \ \  \mathrm{tr}\left\{\boldsymbol{\Omega} \right\}=1,
\end{equation}
where the $4 \times 4$ covariance matrix $\boldsymbol{\Omega}$ is the same for all
antenna pairs and a constraint is required on its trace to remove scaling ambiguities in model (\ref{sirp}). Taking into account such noise model and assuming spatial independence between antenna pairs, the likelihood function is given by
\begin{align}
\label{likelihood}
\nonumber &
f(\mathbf{v}_{12},...,\mathbf{v}_{(M-1)M}|\boldsymbol{\theta },
\boldsymbol{\tau}, \boldsymbol{\Omega})=
\\ &
\prod_{pq}\frac{1}{|\pi
\tau_{pq} \boldsymbol{\Omega}|} \exp
\left\{-\frac{1}{\tau_{pq}}\mathbf{a}_{pq}^{H}(\boldsymbol{\theta})
\boldsymbol{\Omega}^{-1}\mathbf{a}_{pq}(\boldsymbol{\theta})\right\},
\end{align}
with $\boldsymbol{\tau}=[\tau_{12},\tau_{13},\ldots,\tau_{(M-1)M}]^{T}$ and $\mathbf{a}_{pq} = \mathbf{v}_{pq}- \tilde{\mathbf{v}}_{pq}$. In the r.h.s. of (\ref{likelihood}), the product is performed for each antenna pair so there are $B$ elements in the product.
The log-likelihood function is written as follows
\begin{align}
\nonumber
\label{loglikelihood}
& \log f(\mathbf{v}_{12},...,\mathbf{v}_{(M-1)M)}|\boldsymbol{\theta },
\boldsymbol{\tau}, \boldsymbol{\Omega}) =
-4B\log\pi
\\ &
-4\sum_{pq}\log\tau_{pq}-
B\log|\boldsymbol{\Omega}|-\sum_{pq}\frac{1}{\tau_{pq}}\mathbf{a}_{pq}^{H}(\boldsymbol{\theta})
\boldsymbol{\Omega}^{-1}\mathbf{a}_{pq}(\boldsymbol{\theta}).
\end{align}

The proposed robust calibration scheme is based on an iterative ML algorithm \cite{zhang2015maximum, ollier2015article}. The principle is to optimize the log-likelihood function w.r.t.~each unknown parameter, while fixing the others, leading to the so-called concentrated ML estimator. Furthermore, we use in the following a relaxed ML estimator for which the texture parameters are assumed unknown and deterministic. This choice is motivated by the fact that we aim to design a broad robust estimator w.r.t. the presence of outliers but also in order to avoid a misspecification of the probability density function of $\boldsymbol{\tau}$ (i.e., we do not need to specify the texture distribution, ensuring more flexibility). A closed-form expression can be obtained for each texture realization $\tau_{pq}$ and the speckle covariance matrix $\boldsymbol{\Omega}$. Generally, no closed-form expression can be obtained for $\boldsymbol{\theta}$, unless assuming a specific linear modeling of this vector w.r.t.~the noiseless visibilities.

\textbf{\textit{1) Derivation of $\hat{\tau}_{pq}$}}:
We take the derivative of the log-likelihood function in (\ref{loglikelihood}) w.r.t.~$\tau_{pq}$ and
equate it to 0, leading to
\begin{equation}
-\frac{4}{\tau_{pq}}+\frac{1}{\tau_{pq}^{2}}\mathbf{a}_{pq}^{H}(\boldsymbol{\theta})
\boldsymbol{\Omega}^{-1}\mathbf{a}_{pq}(\boldsymbol{\theta})=0.
\end{equation}
We then obtain the expression of the texture estimate,
\begin{equation}
\label{tauExp} \hat{\tau}_{pq} = \frac{1}{4} \mathbf{a}_{pq}^{H}(\boldsymbol{\theta})
\boldsymbol{\Omega}^{-1} \mathbf{a}_{pq}(\boldsymbol{\theta}).
\end{equation}

\textbf{\textit{2) Derivation of $\hat{\boldsymbol{\Omega}}$}}:
We take the derivative of the log-likelihood function w.r.t.~the element
$[\boldsymbol{\Omega}]_{k,l}$ of the speckle covariance matrix  and equate it to 0.
 Using \cite[p. 2741]{hjorungnes2007complex}, we obtain
\begin{equation} 
\label{OmegaDerive}
-B\mathrm{tr}\left\{\boldsymbol{\Omega}^{-1}\mathbf{e}_{k}\mathbf{e}_{l}^{T}\right\}+
\sum_{pq}\frac{1}{\tau_{pq}}\mathbf{a}_{pq}^{H}(\boldsymbol{\theta})
\boldsymbol{\Omega}^{-1}\mathbf{e}_{k}\mathbf{e}_{l}^{T}\boldsymbol{\Omega}^{-1}\mathbf{a}_{pq}(\boldsymbol{\theta})=0
\end{equation}
where the vector $\mathbf{e}_{k}$ contains zeros except at the
$k$-th position which is equal to unity. Using the permutation property of the trace operator, we obtain
\begin{equation}
-B\mathbf{e}_{l}^{T}\boldsymbol{\Omega}^{-1}\mathbf{e}_{k}+
\sum_{pq}\frac{1}{\tau_{pq}}\mathbf{e}_{l}^{T}\boldsymbol{\Omega}^{-1}\mathbf{a}_{pq}(\boldsymbol{\theta})\mathbf{a}_{pq}^{H}(\boldsymbol{\theta})
\boldsymbol{\Omega}^{-1}\mathbf{e}_{k}=0.
\end{equation}
%
Consequently,
\begin{equation}
\label{OmegaEstimBefore}
\hat{\boldsymbol{\Omega}}=\frac{1}{B}\sum_{pq}\frac{1}{\tau_{pq}}\mathbf{a}_{pq}(\boldsymbol{\theta})\mathbf{a}_{pq}^{H}(\boldsymbol{\theta}).
\end{equation}
Since we adopt here an iterative procedure with a concentrated ML scheme, we plug (\ref{tauExp}) into (\ref{OmegaEstimBefore}) leading to
\begin{equation}
\label{OmegaEstim} \hat{\boldsymbol{\Omega}}^{j+1} = \frac{4}{B}
\sum_{pq} \frac{\mathbf{a}_{pq}(\boldsymbol{\theta})
\mathbf{a}_{pq}^{H}(\boldsymbol{\theta})}{\mathbf{a}_{pq}^{H}(\boldsymbol{\theta})(\hat{\boldsymbol{\Omega}}^{j})^{-1}\mathbf{a}_{pq}(\boldsymbol{\theta})}
\end{equation}
where $j$ represents the \textit{j}-th iteration. To ensure uniquely identifiable noise parameters, as it was previously mentioned in (\ref{constraint}), the estimate of $\boldsymbol{\Omega}$ needs to be, e.g., normalized by its trace, as
\begin{equation}
\label{trace}
\hat{\boldsymbol{\Omega}}^{j+1}=\frac{\hat{\boldsymbol{\Omega}}^{j+1}}{\mathrm{tr}\left\{\hat{\boldsymbol{\Omega}}^{j+1}\right\}}.
\end{equation}

\textbf{\textit{3)Estimation of $\hat{\boldsymbol{\theta}}$}}:
Estimating $\hat{\boldsymbol{\theta}}$ for a given $\boldsymbol{\Omega}$ and $\boldsymbol{\tau}$ leads to
\begin{equation}
\label{optimtheta}
\hat{\boldsymbol{\theta }}=\arg \min_{\boldsymbol{\theta } }
\left\{\sum_{pq}\frac{1}{\tau_{pq}}\mathbf{a}_{pq}^{H}(\boldsymbol{\theta})
\boldsymbol{\Omega}^{-1}\mathbf{a}_{pq}(\boldsymbol{\theta})\right\}.
\end{equation}
Depending on the structure of the Jones matrices \cite{noordam1996measurement}, a different procedure can be adopted to estimate $\boldsymbol{\theta}$. A particular parametrization is the non-structured case, where $\boldsymbol{\theta}$ is composed of the entries of all Jones matrices, which is considered in the following.

\section{ESTIMATION OF $\boldsymbol{\hat{\theta}}$ FOR NON-STRUCTURED JONES MATRICES}
\label{sec:nonstructured}

The optimization in (\ref{optimtheta}) may be computationally heavy and very slow in convergence. To overcome this drawback, we apply the expectation-maximization (EM) algorithm, as in \cite{yatawatta2009radio} and \cite{kazemi2011radio}.
Since we adopt the non-structured Jones matrices case (i.e., $\boldsymbol{\theta}$ is a collection of Jones matrices' elements), the vector $\boldsymbol{\theta}$ can be partitioned as
\begin{equation}
\boldsymbol{\theta} = [\boldsymbol{\theta}_{1}^{T}, \ldots,
\boldsymbol{\theta}_{D}^{T}]^{T} =
[\boldsymbol{\theta}_{1_{1}}^{T}, \ldots,
\boldsymbol{\theta}_{1_{M}}^{T}, \ldots,
\boldsymbol{\theta}_{D_{1}}^{T},
\ldots,\boldsymbol{\theta}_{D_{M}}^{T} ]^{T}
\end{equation}
meaning that for the \textit{i}-th source and the \textit{p}-th antenna, we have
$\mathbf{J}_{i_{p}}(\boldsymbol{\theta})=\mathbf{J}_{i_{p}}(\boldsymbol{\theta}_{i_{p}})$ in which $\boldsymbol{\theta}_{i_{p}}$ denotes the parametrization of the path \textit{i}-\textit{p}.

The EM algorithm \cite{dempster1977maximum, mclachlan2007algorithm} is an iterative procedure to approximate the ML estimation technique and reduce its computational cost. First, the E-step computes the conditional expectation of the complete data given the observed data and the current fit for parameters. Second, the M-step maximizes the log-likelihood function of the conditional distribution, previously computed. This may not result in a closed-form expression and requires a numerical optimization procedure. The Levenberg-Marquardt (LM) algorithm is notably particularly appropriate for non-linear problems. The E- and the M-steps are repeated until convergence or until the maximum number of iterations is reached. The complexity is reduced since the unknown parameter vector is partitioned over the sources and optimization is carried out w.r.t.~to $\boldsymbol{\theta}_{i}$ instead of  $\boldsymbol{\theta}$. This leads to single source sub-optimization problems of smaller sizes.

\textbf{\textit{1) E-step}}:
We introduce the complete data vector $\mathbf{w} =
[\mathbf{w}_{1}^{T}, \ldots, \mathbf{w}_{D}^{T}]^{T}$ where, for the
\textit{i}-th source, the  $4B \times 1$  vector $\mathbf{w}_{i}$ is given by
\begin{equation}
\mathbf{w}_{i} =
\mathbf{u}_{i}(\boldsymbol{\theta}_{i})+\mathbf{n}_{i}
\end{equation}
 such that $\mathbf{x}
=\sum_{i=1}^{D}\mathbf{w}_{i}$. The noise vectors $\mathbf{n}_{i}$ are supposed
to be statistically independent such that $\mathbf{n}_{i} \sim \mathcal{CN}(\mathbf{0},\beta_{i}\boldsymbol{\Psi})$ where $\sum_{i=1}^{D}\beta_{i} =
1$ and  $\boldsymbol{\Psi}$ is the covariance matrix of $\mathbf{n}$.
The covariance matrix of each noise vector
$\mathbf{n}_{pq}$ is given by $\tau_{pq} \boldsymbol{\Omega}$.
Making use of the independence property, we obtain $\boldsymbol{\Psi}  =\mathrm{bdiag}\{
    \tau_{12} \boldsymbol{\Omega},
    \ldots, \tau_{(M-1)M}\boldsymbol{\Omega}\}$
%
and the covariance matrix of
$\mathbf{w}$ is given by $\boldsymbol{\Xi}=\mathrm{bdiag}\{
    \beta_{1} \boldsymbol{\Psi}, \ldots, \beta_{D}\boldsymbol{\Psi}\}.$

Using \cite[p.~36]{anderson1958introduction}, we obtain the conditional expectation of the complete data $\hat{\mathbf{w}}=\mathrm{E}\{\mathbf{w}|\mathbf{x};
\boldsymbol{\theta}, \boldsymbol{\tau}, \boldsymbol{\Omega}\}$, in the jointly Gaussian case, that is,
\begin{equation}
\label{wRealEstim} \hat{\mathbf{w}}_{i}
=\mathbf{u}_{i}(\boldsymbol{\theta}_{i}) +
\beta_{i}\left(\mathbf{x}-\sum_{l=1}^{D}\mathbf{u}_{l}(\boldsymbol{\theta}_{l})\right).
\end{equation}

\textbf{\textit{2) M-step}}:
Once $\hat{\mathbf{w}}$ is evaluated, $\boldsymbol{\theta}_{i}$ is estimated through optimization. Independence of $\mathbf{w}_{i}$ leads to
\begin{align}
\nonumber & f(\hat{\mathbf{w}}|\boldsymbol{\theta },\boldsymbol{\tau},\boldsymbol{\Omega}) =
\prod_{i=1}^{D}\frac{1}{|\pi
\beta_{i}\boldsymbol{\Psi}|} 
\\ &
\exp
\left\{-\Big(\hat{\mathbf{w}}_{i}-\mathbf{u}_{i}(\boldsymbol{\theta}_{i})\Big)^{H}(\beta_{i}\boldsymbol{\Psi})^{-1}\Big(\hat{\mathbf{w}}_{i}-\mathbf{u}_{i}(\boldsymbol{\theta}_{i})\Big)\right\}.
\end{align}
For the \textit{i}-th source, the cost function to minimize is given by
$\zeta_{i}(\boldsymbol{\theta}_{i}) = \Big(\hat{\mathbf{w}}_{i}-\mathbf{u}_{i}(\boldsymbol{\theta}_{i})\Big)^{H}(\beta_{i}\boldsymbol{\Psi})^{-1}\Big(\hat{\mathbf{w}}_{i}-\mathbf{u}_{i}(\boldsymbol{\theta}_{i})\Big)$.
At the (\textit{h}+1)-th iteration of the LM-like algorithm, we have:
\begin{equation}
\label{optimLM}
\boldsymbol{\theta}_{i}^{h+1}
=\boldsymbol{\theta}_{i}^{h}-(\nabla_{\boldsymbol{\theta}_{i}}\nabla_{\boldsymbol{\theta}_{i}}^{T}
\zeta_{i}(\boldsymbol{\theta}_{i})+\lambda\mathbf{I}_{4M})^{-1}\nabla_{\boldsymbol{\theta}_{i}}
\zeta_{i}(\boldsymbol{\theta}_{i})| _{\boldsymbol{\theta}_{i}^{h}}.
\end{equation}

\LinesNumberedHidden
 \begin{algorithm}
\SetAlgorithmName{Proposed algorithm}{}{} \caption{}
\SetKwInOut{input}{input} \SetKwInOut{output}{output}
\SetKwInOut{initialize}{initialize}
\input{$D$, $M$, $B$, $\mathbf{C}_{i}$,  $\beta_{i}$, $\mathbf{x}$}
\output{estimate of $\boldsymbol{\theta}$}
\initialize{$\hat{\boldsymbol{\Omega}}$ $\leftarrow$ $\boldsymbol{\Omega}_{\mathrm{init}}$,$\hat{\boldsymbol{\tau}}$ $\leftarrow$ $\boldsymbol{\tau}_{\mathrm{init}}$}
\While
{stop criterion unreached}
{\setcounter{AlgoLine}{0} \ShowLn
$\hat{\boldsymbol{\theta}}_{i}\leftarrow\boldsymbol{\theta}_{i_{\mathrm{init}}}$, $i=1, \hdots, D$\\
\While
{stop criterion unreached}
{\setcounter{AlgoLine}{1}
\ShowLn E-step: $\hat{\mathbf{w}}_{i}$ obtained with (\ref{wRealEstim})\\
\ShowLn M-step: $\hat{\boldsymbol{\theta}}_{i}$ obtained with (\ref{optimLM})\\
}\setcounter{AlgoLine}{3}
\ShowLn Obtain $\hat{\boldsymbol{\Omega}}$ with (\ref{OmegaEstim}) then (\ref{trace})\\
\ShowLn Obtain $\hat{\boldsymbol{\tau}}$ with (\ref{tauExp})\\
}
\end{algorithm}

\section{NUMERICAL SIMULATIONS}
\label{sec:Simus}

\begin{figure}[t] 
  \centering
  \centerline{\includegraphics[width=8.5cm]{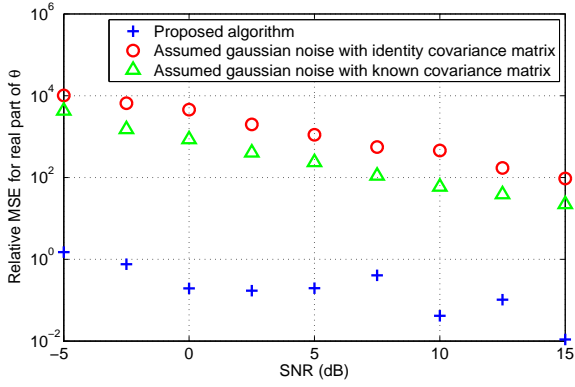}}
\caption{Evolution of the relative MSE of a given unknown parameter as a function of the SNR.} \label{fig:res1}
\end{figure}

\begin{figure}[t] 
  \centering
  \centerline{\includegraphics[width=8.5cm]{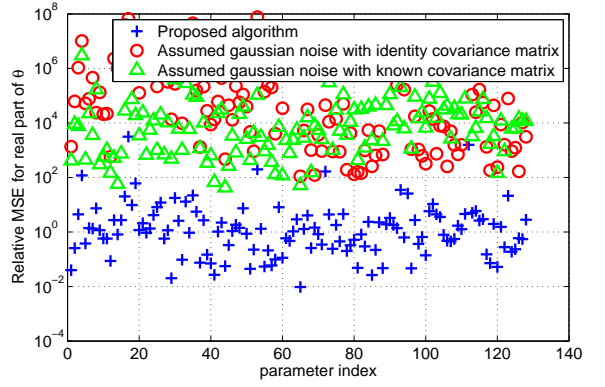}}
\caption{Relative MSE of the 128 unknown parameters for a given SNR.} \label{fig:res2}
\end{figure}

In the following simulations, we consider $D=2$ sources and $M=16$ antennas. Therefore, the number of antenna pairs is $B=120$, each providing a $4 \times 1$ observation vector, and there are 128 unknown parameters to estimate, corresponding to the entries of the Jones matrices. The source coherency $\mathbf{C}_{i}$ is defined thanks to the Stokes parameters \cite{hamaker1996understanding, smirnov2011revisiting}, which represent the polarization state of the \textit{i}-th signal source considered. The number of Monte Carlo runs is set to 100.

In Fig.~\ref{fig:res1}, we plot the relative mean square error (MSE) vs.~ (residual) signal-to-noise ratio (SNR), for a given parameter representative of the overall behavior (by residual SNR, we mean the SNR computed using only the off-diagonal terms of the covariance matrix since the diagonal terms have been deleted, i.e., equivalently $\tilde{\mathbf{V}}_{pp}$ is not considered in the observation model (\ref{ME})). We represent the following cases: i) the proposed algorithm as exposed in the pattern (blue curve) which intends to propose a robust estimator, ii) the case when gaussian noise is assumed with a known covariance matrix (equivalently $\boldsymbol{\Omega}$ is known and $\boldsymbol{\tau}$ is set to a vector filled with ones during the whole estimation procedure (green curve)) and iii) the case when gaussian noise is assumed with an identity covariance matrix (equivalently $\boldsymbol{\Omega}$ is set to the identity matrix and $\boldsymbol{\tau}$ is filled with ones (red curve)). The two last cases correspond to Gaussian modeling, with spatially correlated and i.i.d. noise. For all cases, the observations are generated using the true noise covariance matrix, structured as described in (\ref{sirp}).
In Fig.~\ref{fig:res2}, we plot the relative MSE of each unknown parameter for a SNR of 15 dB.

The lowest MSE is achieved with the proposed algorithm, which estimates iteratively both texture and speckle components and the Jones matrices for calibration. Such performance is due to the SIRP noise assumption which includes many various distributions. The calculations were performed without precising the distribution considered, thus ensuring robust calibration to the presence of outliers in the noise (faint sources). If we do not characterize precisely the noise (i.e., we dot not take into account the probability density function of $\boldsymbol{\tau}$), we have less information (relaxed ML estimator) but we reach robustness. The two other curves do not take into account the noise model as presented in (\ref{sirp}), leading to misspecifications and poor accuracy in calibration.

\section{CONCLUSION}
\label{sec:Conclusion}

In this paper, we propose a robust calibration algorithm of radio interferometers, where non-structured Jones matrices are used to model the different perturbations introduced along the signal path. The derived scheme is based on an iterative relaxed concentrated ML method, in which a SIRP noise model is introduced, without fixing the texture distribution. Closed-form expressions are obtained for the noise parameters (texture components and speckle covariance matrix) while the estimation of the unknown vector $\boldsymbol{\theta}$ is performed via an optimization process. The computational complexity of the problem is reduced with the use of the EM algorithm and the partition per source.

\bibliographystyle{IEEEbib}
\bibliography{nab}

\end{document}